\documentclass{appolb}
\usepackage{amsmath}
\usepackage{amssymb}
\usepackage{psfrag,epsfig,graphicx,graphics}
\usepackage{caption}
\usepackage{multirow}
\usepackage{verbatim}
\usepackage{color}
\usepackage{xcolor}

\def\meson{m}

\def\bea{\begin{eqnarray}}
	\def\eea{\end{eqnarray}}

\newcommand{\Photo}{}

\usepackage[utf8]{inputenc}
\def\lsim{\raise0.3ex\hbox{$<$\kern-0.75em\raise-1.1ex\hbox{$\sim$}}}
\def\gsim{\raise0.3ex\hbox{$>$\kern-0.75em\raise-1.1ex\hbox{$\sim$}}}

\def\bei{\begin{itemize}}
	\def\ei{\end{itemize}}
\def\bea{\begin{eqnarray}}
	\def\eea{\end{eqnarray}}
\def\beas{\begin{eqnarray*}}
	\def\eeas{\end{eqnarray*}}
\def\beqas{\begin{eqnarray*}}
	\def\eqas{\end{eqnarray*}}
\def\beq{\begin{equation}}
	\def\eeq{\end{equation}}
\def\beqd{\begin{displaymath}}
	\def\eeqd{\end{displaymath}}
\def\eqd{\end{displaymath}}

\def\beeq{\begin{eqnarray}} \def\eeeq{\end{eqnarray}}

\def\bef{\begin{frame}}

\def\slashchar#1{\setbox0=\hbox{$#1$}
\dimen0=\wd0
\setbox1=\hbox{/} \dimen1=\wd1
\ifdim\dimen0>\dimen1
\rlap{\hbox to \dimen0{\hfil/\hfil}}
#1
\else
\rlap{\hbox to \dimen1{\hfil$#1$\hfil}}
/
\fi}

\newcommand{\be}{\begin{equation}}
\newcommand{\ee}{\end{equation}}
\newcommand{\eq}{\end{equation}}

\newcommand{\fin}{\end{document}}

\newcommand{\com}[1]{\textcolor{orange}{#1}}

\def\pv{\vec{p}_t}
\def\dv{\vec{\Delta}_t}

\def\beqa{\begin{eqnarray}}
\def\eqa{\end{eqnarray}}

\begin{document}
\title{Accessing GPDs through the exclusive photoproduction of a photon-meson pair with a large invariant mass
\thanks{Presented by S.~Nabeebaccus at ``Diffraction and Low-$x$ 2022'', Corigliano Calabro (Italy), September 24-30, 2022.}
}
\author{Goran Duplan\v{c}i\'{c}$ ^{1} $, Saad Nabeebaccus$ ^{2} $, Kornelija Passek-Kumeri\v{c}ki$ ^{1} $, Bernard Pire$ ^{3} $, Lech Szymanowski$ ^{4} $ and Samuel Wallon$ ^{2} $
\address{\vspace{0.4cm}$^1$Theoretical Physics Division, Rudjer Bo{\v s}kovi{\'c} Institute,
	HR-10002 Zagreb, Croatia\\
	$^2$Universit\'e Paris-Saclay, CNRS/IN2P3, IJCLab, 91405 Orsay, France\\  
	$^3$CPHT, CNRS, Ecole polytechnique, Institut Polytechnique de Paris, 91128 Palaiseau, France\\  
	$^4$National Centre for Nuclear Research (NCBJ), Warsaw, Poland }
}
\maketitle
\begin{abstract}
\vspace{0.1cm}We study the exclusive photoproduction of a photon-meson pair with a large invariant mass, working in the QCD factorisation framework. Explicitly, we consider a $ \rho $-meson or a charged $  \pi  $ in the final state. This process gives access to chiral-even GPDs as well as chiral-odd GPDs. We focus here on the chiral-even sector. The computation is performed at leading order and leading twist. We discuss the prospects of measuring them in various experiments such as JLab 12-GeV, COMPASS, future EIC and LHC (in ultraperipheral collisions). In particular, the high centre of mass energies available at collider experiments can be used to probe GPDs at small skewness $ \xi $. We also compute the polarisation asymmetries with respect to the incoming photon. The results for an alternative distribution amplitude (`holographic' form) are also compared with the predictions obtained with an asymptotic distribution amplitude.
\end{abstract}
  
\section{Introduction}

A new family of 2 → 3
exclusive processes~\cite{Ivanov:2002jj,ElBeiyad:2010pji,Pedrak:2017cpp,Pire:2019hos,Pedrak:2020mfm,Cosyn:2021dyr} has been shown to be very promising in view of accessing generalised parton distributions (GPDs).
In the present work, we focus on the exclusive photoproduction of a photon-meson pair with a large invariant mass. Work in this direction has already been performed in \cite{Boussarie:2016qop} for the case of a $ \rho $-meson in the final state, and in \cite{Duplancic:2018bum,Duplancic:2022ffo} for a charged pion in the final state. Imposing a large value for the invariant mass of the photon-meson pair provides the hard scale for employing collinear QCD factorisation.
Recently, QCD factorisation has been proved for a family of exclusive $ 2 \to 3 $ processes \cite{Qiu:2022bpq,Qiu:2022pla} at the leading twist, which includes the process we study. The proof of factorisation relies on the transverse momentum of the outgoing photon/meson to be large, rather than their invariant mass, which is a stricter condition.

On the one hand, one of the main advantages of studying this channel is that for a transversely polarised $  \rho  $ meson, this process gives access to chiral-odd GPDs at \textit{leading twist}, unlike in deeply virtual meson production (DVMP). Since chiral-odd GPDs are not well-known experimentally, this provides an excellent opportunity to study them. On the other hand, these new channels with 3 particles in the final state offer complementary ways to access the chiral-even sector of GPD, besides the deeply virtual Compton scattering and DVMP. We presently focus on the chiral-even sector, which we illustrate with the case of a charged pion.

More specifically, the process we study is
\begin{align}
	\gamma (q)+N(p_1) \longrightarrow \gamma (k)+N'(p_2)+m(p_{\meson})\;,
\end{align}
where $ \meson= \rho^{0,\pm}_{L,T},  \pi ^{\pm} $.
We denote the masses of the nucleon and the meson to be $M$ and $M_\meson$  respectively.
The use of collinear QCD factorisation requires that $ -u'= \left( p_{\meson}-q \right)^2  $, $ -t'= \left( k-q \right)^2  $ and $ M_{\gamma \meson}^2 =  \left(  p_{\meson}+k\right)^2  $ to be large, while $ -t =  \left( p_2-p_1 \right)^2  $ needs to be small. For this, we employ the cuts $ -u',-t'>1$ GeV$ ^2 $, and $ -t < 0.5 $ GeV$ ^2 $. We note that these cuts are sufficient to ensure that $ M^2_{\gamma\meson} > 1 $ GeV$ ^2 $.
More details regarding the kinematics can be found in \cite{Boussarie:2016qop,Duplancic:2018bum}.
The results will be expressed as functions of  $ \left( -u' \right),\,(-t)$, and $ M_{\gamma \meson}^2 $.

\section{Computation}

The chiral-even light-cone distribution amplitude (DA), e.g. for the $\pi^+$ meson is defined, at the leading 
twist 2, by the matrix element, 
\begin{equation}
	\langle \pi^{+}(p_\pi)|\bar{u}(y)\gamma^5 \gamma^\mu  d(-y)|0 \rangle = i f_{\pi} 
	p_\pi^\mu \int_0^1dz\ e^{-i(z - \bar{z}) p_\pi \cdot y}\ \phi_{\pi}(z),
	\label{defDApi}
\end{equation}
with the decay constant 
 $f_{\pi}=131\,\mbox{MeV}$.  For the computation, we use the asymptotic form of the DA, as well as an alternative form, which we call `holographic' DA, both normalised to 1, given by
\beqa
\label{DA-asymp}
\phi^{\rm as}(z)= 6 z (1-z)\;,\quad \phi^{\rm hol}(z)= \frac{8}{ \pi } \sqrt{z (1-z)}\;.
\eqa

The chiral-even vector GPDs of a quark $q$ (where $q = u,\ d$) in the nucleon target   are  defined by
\beqa
&&\langle p(p_2,\lambda')|\, \bar{q}\left(-\frac{y}{2}\right)\,\gamma^+q \left(\frac{y}{2}\right)|p(p_1,\lambda) \rangle = \int_{-1}^1dx\ e^{-\frac{i}{2}x(p_1^++p_2^+)y^-} \\ \nonumber 
&&\times \bar{u}(p_2,\lambda')\, \left[ \gamma^+ H^{q}(x,\xi,t)   +\frac{i}{2m}\sigma^{+ \,\alpha}\Delta_\alpha  \,E^{q}(x,\xi,t) \right]
u(p_1,\lambda)\,,
\label{defGPDEvenV}
\eqa
and analogously for chiral-even axial GPDs. In our analysis, 
the contributions from the the chiral-even axial GPD $ E^{q} $ and $  \tilde{E}^{q}  $  are neglected, since they are suppressed by kinematical factors at the cross-section level. The GPDs are parametrised through double distributions. We note that for the modelling of the chiral-even axial GPDs, we use two different parametrisations for the input PDFs: The \textit{standard} scenario, for which the light sea quark and anti-quark distributions are \textit{flavour-symmetric}, and the \textit{valence} scenario which corresponds to completely \textit{flavour-asymmetric} light sea quark densities. More details can be found in \cite{Boussarie:2016qop,Duplancic:2018bum}.

The amplitude for the process is expressed as the convolution over $ x $ and $ z $ of the coefficient function (hard part), the GPD and the DA. 
The fully differential cross-section, as a function of $ -u' $, $ -t $ and $ M_{\gamma \meson}^2 $,  is then given by
\begin{equation}
	\label{difcrosec}
	\left.\frac{d\sigma}{dt \,du' \, dM^2_{\gamma\meson}}\right|_{\ -t=(-t)_{ \mathrm{min} }} = \frac{|\mathcal{\overline{M}}|^2}{32S_{\gamma N}^2M^2_{\gamma\meson}(2\pi)^3}\;,
\end{equation}
where $ -t $ has been set to the minimum value $ (-t)_{ \mathrm{min} } $ allowed by the kinematics, including the imposed cuts, and is in general a function of $ M^2_{\gamma \meson} $ and $ S_{\gamma N} $. We refer the reader to \cite{Boussarie:2016qop,Duplancic:2018bum,Duplancic:2022ffo} for the details regarding the computation, the integration over the phase space and the computation of the linear polarisation asymmetry (LPA) wrt to the incoming photon. Note that the circular asymmetry vanishes for the present unpolarised target case.

\section{Results}

\begin{figure}[h]
	\vspace{0.3cm}
	
	\psfrag{HHH}{\hspace{-1.5cm}\raisebox{-.6cm}{\scalebox{.8}{$-u' ({\rm 
					GeV}^{2})$}}}
	\psfrag{VVV}{\raisebox{.3cm}{\scalebox{.8}{$\hspace{-.1cm}\displaystyle\left.\frac{d 
					\sigma_{\gamma \pi^+}}{d M^2_{\gamma \pi^+} d(-u') d(-t)}\right|_{(-t)_{\rm min}}({\rm pb} \cdot {\rm GeV}^{-6})$}}}
	\psfrag{TTT}{}

	\hspace{0.55cm}\scalebox{.875}{\centerline{	\hspace{0.4cm}	{\includegraphics[width=17pc]{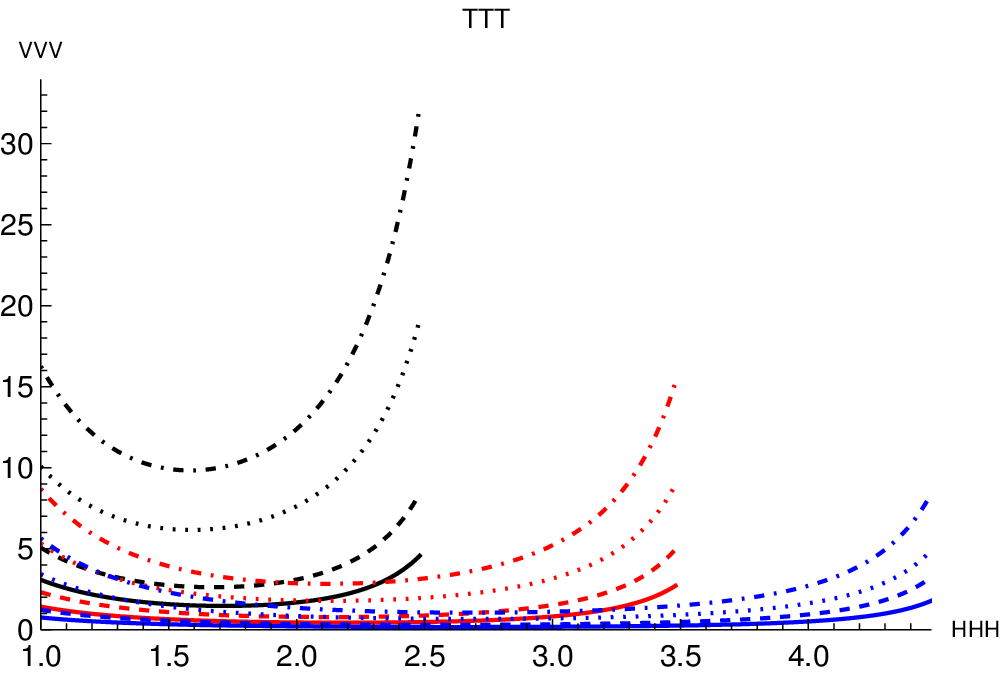}}
			\psfrag{HHH}{\hspace{-1.5cm}\raisebox{-.6cm}{\scalebox{.8}{$M^{2}_{\gamma  \pi^+ } ({\rm 
						GeV}^{2})$}}}
		\psfrag{VVV}{\raisebox{.3cm}{\scalebox{.8}{$\hspace{-.4cm}\displaystyle\frac{d 
						\sigma_{\gamma\pi^+}}{d M^2_{\gamma \pi^+} }({\rm pb} \cdot {\rm GeV}^{-2})$}}}
		\psfrag{TTT}{}
				\includegraphics[width=17pc]{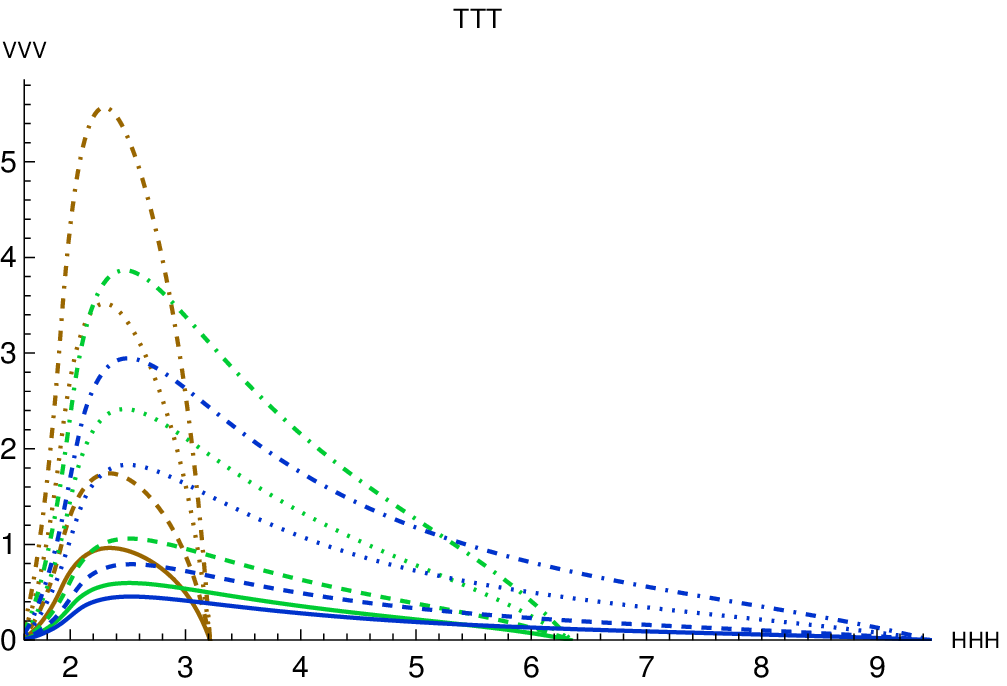}}}
	
	\vspace{0.4cm}
	
	\caption{\small Left: The fully differential cross-section for $  \pi^+ $ as a function of $ -u' $ is shown. $ M_{\gamma  \pi }^2 =3,4,5$ GeV$ ^2 $ correspond to black, red and blue respectively. The difference between standard (dotted) and valence (solid) scenarios for an asymptotical DA, and between standard (dot-dashed) and valence (dashed) scenarios for a holographic DA is also illustrated.  $ S_{\gamma N}$ is fixed at 20 GeV$ ^2 $. Right: The single differential cross-section for $  \pi^+ $ as a function of $ M_{\gamma  \pi^+}^2 $. The values $ S_{\gamma N} =8,\,14,\,20$ GeV$ ^2 $ correspond to brown, green and blue respectively. The same line style conventions for the GPD and DA models are used for both plots.}
	\label{Fig:diffXsection}
\end{figure}

We present only a few plots which are representative, focusing on the  $ \pi^{+} $ case. In Fig.~\ref{Fig:diffXsection} on the left, we show the fully differential rate as a function of $ -u' $, for different values of $ M_{\gamma \pi^+}^2 $. The effect of using the two different models for the distribution amplitude, as well as that of using the valence and standard scenarios for modelling the GPDs, is also illustrated. We thus find that using the holographic DA gives a result that is roughly twice that of the asymptotical DA. Still, to properly distinguish between the two models, one would need to include NLO corrections, since they can be large.

The single differential cross-section as a function of $ M_{\gamma  \pi^+ } ^2$ for different values of $ S_{\gamma N} $ is shown on the right plot in Fig.~\ref{Fig:diffXsection}. We note that while the fully differential cross-section is largest for smaller $ M_{\gamma  \pi^+ }^2 $, the range of $ -u' $ is more restricted, due to the shrinking of the phase space. In fact, there is a compromise between the two effects, and this explains the position of the peak around $ M_{\gamma  \pi^+ }^2 \approx 3 $ GeV$ ^2 $ in the single differential cross-section plot on the right of Fig.~\ref{Fig:diffXsection}. More complete results, see~\cite{Duplancic:2022ffo}, show that the position of this peak is more or less the same as $ S_{\gamma N} $ increases beyond $ 20  $ GeV$ ^2 $.

\begin{figure}[h!]
	
	\psfrag{HHH}{\hspace{-1.5cm}\raisebox{-.6cm}{\scalebox{.8}{$S_{\gamma N} ({\rm 
					GeV}^{2})$}}}
	\psfrag{VVV}{\raisebox{.3cm}{\scalebox{.9}{$\hspace{-.4cm}\displaystyle
				\sigma_{\gamma\pi^+}({\rm pb} )$}}}
	\psfrag{TTT}{}

	\hspace{.8cm}\scalebox{0.87}{\centerline{
		{\includegraphics[width=17pc]{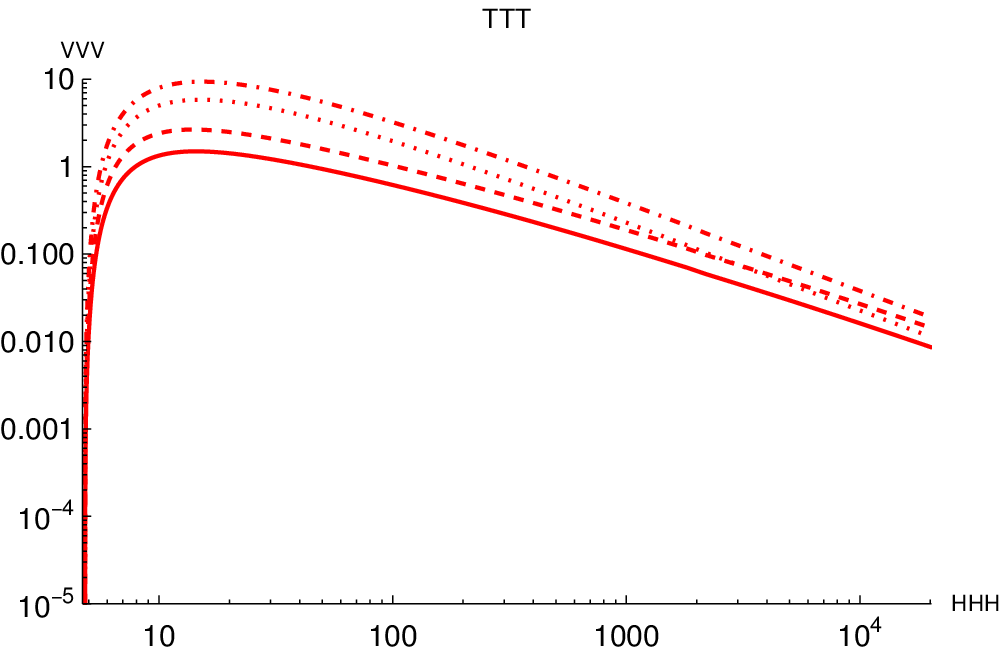}}
			\psfrag{HHH}{\hspace{-1.5cm}\raisebox{-.6cm}{\scalebox{.8}{$M^{2}_{\gamma  \pi^+ } ({\rm 
						GeV}^{2})$}}}
		\psfrag{VVV}{LPA}
		\psfrag{TTT}{}
		{\includegraphics[width=17pc]{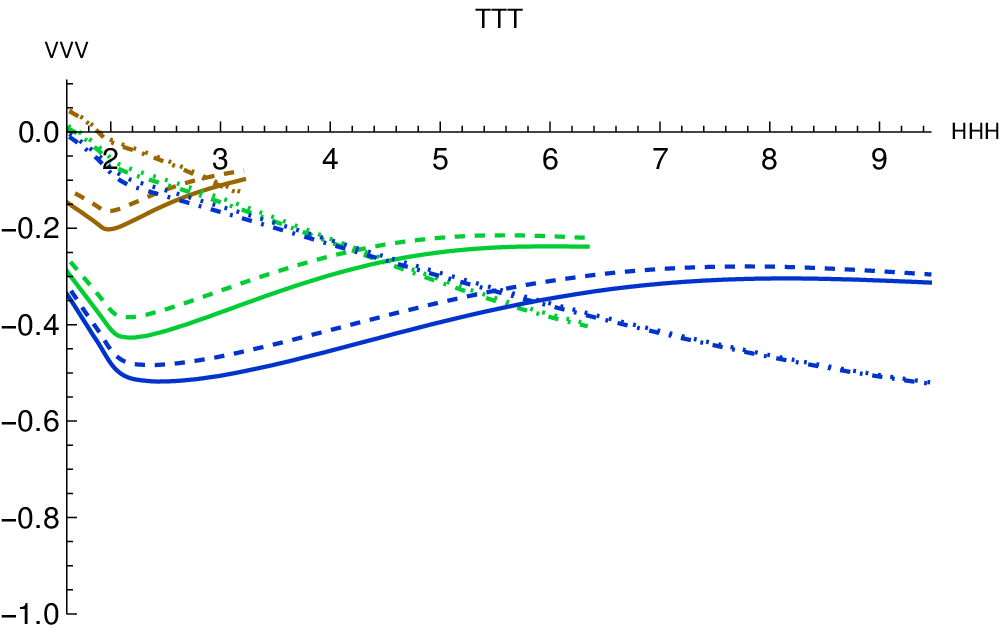}}}}
	
	\vspace{0.2cm}
	\caption{\small Left: The plot shows the cross-section $  \sigma _{\gamma  \pi^+} $ as a function of the centre of mass energy $ S_{\gamma N} $. Right: The LPA wrt the incoming photon for the $ \pi^+ $ case is shown as a function of $ M_{\gamma  \pi ^+}^2 $, at the single differential level. $ S_{\gamma N} =$ 8, 14, 20 GeV$ ^2 $ correspond to brown, green and blue respectively. In both plots, the same line style conventions as in Fig.~\ref{Fig:diffXsection} are used.}
	\label{Fig:SigmavsSgN}
\end{figure}

Finally, in Fig.~\ref{Fig:SigmavsSgN}, we show the variation of the cross-section as a function of $ S_{\gamma N} $ (left). The cross-section drops rather rapidly with $ S_{\gamma N} $, and has a peak at around 20 GeV$ ^2 $ (note the log scales for both axes). We note that while LHC can access very high energies, the photon flux from the Pb nucleus in p-Pb collisions decreases very rapidly with $ S_{\gamma N} $. This, coupled with the fact the cross-section itself decreases with increasing $ S_{\gamma N} $, implies that the total cross-section is dominated by the region of relatively small $ S_{\gamma N} $. The plot on the right of Fig.~\ref{Fig:SigmavsSgN} corresponds to the LPA at the single differential level, as a function of $ M_{\gamma  \pi ^{+}}^2 $. An interesting feature of the plot is that the shape of the curves is very different for the two GPD models we consider, and therefore, the LPA could be used to distinguish them.

The counting rates for $ \rho _L^0 $, $ \rho _L^+ $ and $ \pi^+ $ mesons for LHC in UPC and future EIC are shown in Table \ref{tab:counting-rates}. For LHC, we used an integrated luminosity of $ 1200 $ nb$ ^{-1} $, while for EIC, we used an expected integrated luminosity of $ 10^7 $ nb$ ^{-1} $. The range for the counting rates in each case is obtained by considering the minimum and maximum obtained from the different models (holographic DA vs asymptotical DA, and valence vs standard scenarios). Two sets of counting rates are shown, one without any cut in $ S_{\gamma N} $ and the other with a cut of $ S_{\gamma N} \geq 300 $ $ \mathrm{GeV}^2  $. Introducing a lower bound on $ S_{\gamma N} $ allows us to study GPDs in the small $  \xi  $ region. At $ S_{\gamma N} =300 $ $  \mathrm{GeV}^2 $, we find that the region of $ M_{\gamma \meson}^2  $ where the cross-section is maximum (see Fig.~\ref{Fig:diffXsection}) corresponds to $  \xi  \approx 5  \cdot 10^{-3} $, and it goes down to $  \xi \approx 7.5 \times 10^{-5}$ at $ S_{\gamma N}=20000 $ $ \mathrm{GeV}^2  $. Despite the fact that the number of events is dominated by the region of $ S_{\gamma N} \leq 300 $ $ \mathrm{GeV}^2 $, we find that there may still be reasonable statistics to prompt a study of our process in the small $  \xi  $ region at LHC and EIC.

\renewcommand{\arraystretch}{1.1}

\begin{table}[h!]
	\centering
	\begin{tabular}{ |c|c |c|c|}
		\hline
	Experiment &	Meson & 		Without cut & $S_{\gamma N} \geq 300 $ $ \mathrm{GeV}^2  $\\
		\hline\hline
	\multirow{3}{*}{LHC in UPC} &	$  \rho _L^0 $ &  8.7--16 $ \times  10^3$ & 4.1--8.1 $ \times 10^2 $ \\
		\cline{2-4}
	&	$  \rho _L^+  $ & 4.8--11 $ \times  10^3$& 2.1--6.4 $ \times 10^2 $\\		\cline{2-4}
	&	$  \pi ^+ $ & 1.6--9.3 $ \times  10^3$& 1.0--3.4 $ \times 10^2 $\\
		\hline 
			\multirow{3}{*}{future EIC} &	$  \rho _L^0 $ & 13--24 $ \times  10^3$ & 5.9--12 $ \times 10^2 $ \\
		\cline{2-4}
		&	$  \rho _L^+  $ &  7.0--15$ \times  10^3$& 3.1--9.3 $ \times 10^2 $\\
		\cline{2-4}
		&	$  \pi ^+ $ &  2.3--13$ \times  10^3$& 1.4--5.0 $ \times 10^2 $\\
		\hline 
	\end{tabular}
	\caption{\small The counting rates for $ \rho _L^0 $, $ \rho _L^+ $ and $ \pi^+ $ mesons for LHC in UPC and future EIC are shown. The third column shows the counting rates without any cuts in $ S_{\gamma N} $, while the fourth corresponds to having a cut of $ S_{\gamma N} \geq 300$ $  \mathrm{GeV}^2  $, which gives access to the small $  \xi  $ region. }
	\label{tab:counting-rates}
\end{table}

The counting rates for the JLab 12-GeV experiment, which are roughly one order of magnitude larger than those reported in the third column of Table \ref{tab:counting-rates}, can be found in \cite{Duplancic:2022ffo}. Although the statistics are lower for p-Pb UPCs at LHC and EIC, the energies that can be accessed are higher. This may enable a study of GPDs at small skewness $ \xi $ to be performed.

\bibliographystyle{JHEP}

\bibliography{/home/saadn/work/projects/common/masterrefs}

\end{document}